# High-entropy oxide photocatalysts for green ammonia synthesis from nitrogen fixation in water

Thanh Tam Nguyen[a,b], Jacqueline Hidalgo-Jiménez[a,c] and Kaveh Edalati[a,b,]*

[a] WPI, International Institute for Carbon Neutral Energy Research (WPI-I2CNER), Kyushu University, Fukuoka 819-0395, Japan
[b] Mitsui Chemicals, Inc. - Carbon Neutral Research Center (MCI-CNRC), Kyushu University, Fukuoka 819-0395, Japan
[c] Graduate School of Integrated Frontier Sciences, Department of Automotive Science, Kyushu University, Fukuoka 819-0395, Japan

Ammonia, a critical chemical fertilizer and a potential hydrogen carrier, can be sustainably synthesized from atmospheric nitrogen and water under ambient conditions through photocatalysis. In this study, high-entropy oxides with $d^0$ and mixed $d^0+d^{10}$ cationic configurations are introduced as a new group of catalysts for nitrogen fixation and photocatalytic ammonia production. The oxides exhibit impressive efficiency in ammonia production compared to binary oxides, while the efficiency is improved by using a mixed cationic configuration. It was shown that the incorporation of $d^{10}$ elements, such as gallium and zinc, boosts the photocatalytic reactions by improving light absorbance, charge separation and charge lifetime. These findings demonstrate the potential of high-entropy oxides as next-generation photocatalysts for green ammonia synthesis, offering an effective alternative to conventional catalytic systems.

*Corresponding author (E-mail: kaveh.edalati@kyudai.jp; Phone: +80 92 802 6744)

Nitrogen ($N_2$), constituting approximately 78% of the Earth's atmosphere, is an essential element for the synthesis of chemical and biological compounds such as ammonia ($NH_3$), amino acids and nucleotides [1]. Ammonia, serving as a critical precursor for the synthesis of fertilizers, explosives and other chemicals, is naturally reduced from atmospheric nitrogen by diazotrophs and contributes to the biological nitrogen cycle on the Earth [2]. However, the activity of diazotrophs is insufficient to satisfy the demands of the fertilizer industry. In addition to chemical and biological applications, ammonia is emerging as a prominent hydrogen storage and transport medium because of its condensation ability to form a liquid. This positions ammonia as a potential key player in future hydrogen economies.

Currently, the predominant ammonia production technology is the Haber-Bosch process, which reacts nitrogen with hydrogen over Fe- or Ru-based catalysts at high temperature (573-773 K) and pressure (10-20 MPa) to overcome the high N-N triple bond dissociation energy (941 kJ $mol^{-1}$) [3]. As a result, this energy-intensive Haber-Bosch method utilizes around 2% of global anthropogenic energy and releases approximately 400 Mt of $CO_2$ annually, contributing to 1.6% of total worldwide $CO_2$ emissions [4]. Environmentally friendly and low-energy solutions for ammonia synthesis are gaining attention because of the pressing need to decrease energy consumption and greenhouse gas emissions.

Photocatalytic ammonia production has emerged as a promising approach due to its ability to harness solar energy, a clean, abundant and renewable resource, to drive the transformation of nitrogen and water into ammonia under eco-friendly conditions [4-7]. Since the first report of photocatalytic ammonia production using $TiO_2$ in 1977 [8], numerous semiconductor materials have been explored for this application. Some main groups of photocatalysts for ammonia production are metal oxides ($TiO_2$ [9], $Fe_2O_3$ [10] and $Cu_2O \cdot xH_2O$ [11]), metal sulfides (CdS [12], FeMoS [13], $Zn_{0.1}Sn_{0.1}Cd_{0.8}S$ [14] and $MoS_2$ [15]), bismuth oxyhalides (BiOBr [16,17], $Bi_5O_7Br$ [18,19] and $Bi_5O_7I$ [20]) and carbonaceous materials (boron-doped diamond [21], metal-diamond heterostructures [22] and $g-C_3N_4$ [23]). Despite notable progress, these materials suffer from some limitations, including poor light absorbance, high recombination rates of photoexcited charge carriers, limited active sites and insufficient stability, all of which hinder their practical application in ammonia production. Therefore, developing efficient, stable and durable photocatalysts remains crucial for achieving economically viable ammonia synthesis by photocatalysis.

High-entropy ceramics, a novel class of materials, have recently drawn attention for their outstanding stability and multifunctional properties in various catalytic and energy-related applications. These materials are, in general, described as multi-component systems composed of five or more principal cations, with a high mixing entropy exceeding 1.5$R$ ($R$: the gas constant). Their exceptional stability stems from their low Gibbs free energy (resulting from high entropy), whereas their exclusive physical and chemical properties arise from a combination of lattice strain, heterogeneous distribution of electrons and the cocktail effect [24]. High-entropy oxides (HEOs) have the highest popularity among high-entropy ceramic materials, being used for various photocatalytic applications, particularly by the group of current authors. These applications include water splitting [25,26], $CO_2$ conversion [27,28] and plastic waste transformation [29,30]. Despite the reported high efficiency of HEOs for catalysis, no prior studies have explored their potential for photocatalytic ammonia synthesis.

This work pioneers the investigation of high-entropy photocatalysts for ammonia production. Two groups of HEOs are examined: (i) systems incorporating elements with only a $d^0$ cationic configuration (Ti, Zr, Nb, Ta and Hf), which are widely recognized as effective photocatalysts [31], and (ii) mixed systems combining $d^0$ cations (Ti, Zr, Nb and Ta, known for their strong electron-accepting ability) and $d^{10}$ cations (Zn and Ga, with the capability to serve as electron donors). The findings reveal that high-entropy materials exhibit higher photocatalytic activity for ammonia production compared to binary oxides, unveiling a novel photocatalytic function for this emerging class of materials. Moreover, a comparative analysis of HEOs with purely $d^0$ cations versus mixed $d^0$-$d^{10}$ compositions underscores the enhancement in photocatalytic performance achieved through the tailored electronic configuration of mixed cations.

The T-Zr-Nb-Ta-X-based HEOs (X: Hf, Ga and Zn) were synthesized from pure elements or binary oxides following a procedure described in earlier studies [26,28,29]. To synthesize the Hf-containing oxide, elements with >99.7% purity with equal atomic fractions and a 10 g mass were arc-melted, flipped and re-melted seven times under pure argon using a crucible of water-cooled copper and an electrode of tungsten. The ingot was then cut to discs (0.1 cm diameter and 0.08 cm thickness) and mechanically homogenized by high-pressure torsion at ambient temperature, 6 GPa pressure, 10 turns and 1 rpm rotation speed. The homogenized discs were then air-oxidized at 1373 K for 48 h. The Ga- and Zn-containing oxides were synthesized from corresponding binary oxides with >97% purity. The binary oxide powders were mixed in acetone by mortar and pestle for 0.5 h, compressed into 10 mm diameter discs, processed by high-pressure torsion at room temperature, 6 GPa pressure, 6 turns and 1 rpm rotation speed, and air-calcinated for 24 h at 1373 K. To enhance homogeneity, the calcinated powders were further treated by high-pressure torsion for another three turns, and re-calcinated at 1373 K for 24 h.

The crystal structures of the HEOs were investigated by (i) X-ray diffraction (XRD using Cu K$\alpha$), (ii) Raman measurement ($\lambda$ = 532 nm laser light), and (iii) transmission electron microscopy (TEM using 200 keV electrons) with high-resolution micrographs and corresponding fast Fourier transform (FFT). The morphology of powders and elemental distribution were analyzed by scanning electron microscopy (SEM using 15 keV electrons) coupled with energy-dispersive X-ray spectroscopy (EDS). The surface areas of the HEOs were identified by either particle size measurement via SEM or the Brunauer-Emmett-Teller (BET) method and nitrogen adsorption. Optical characteristics were investigated using a UV-Vis diffuse reflectance spectrometer and the Kubelka-Munk bandgap measurement analysis. The valence band maximum was estimated by X-ray photoelectron spectroscopy (XPS with Al K$\alpha$ irradiation). The charge carrier recombination behavior was investigated using (i) photoluminescence spectroscopy (PL with a laser lamp of 325 nm wavelength), and (ii) time-resolved photoluminescence (PL decay using a laser lamp of 340 nm wavelength and a 385 nm long-pass filter). Photocurrent measurements were carried out in 1 M KOH electrolyte using a three-electrode configuration. The catalyst-coated FTO glass served as the working electrode, a Pt wire was used as the counter electrode, and an Ag/AgCl electrode functioned as the reference electrode. Potentiostatic amperometry was performed under chopped light illumination, with cycles consisting of 60 s of light exposure followed by 60 s of darkness, repeated across multiple measurements. The oxygen vacancy behavior was investigated by electron paramagnetic resonance (EPR) analysis by a 9.4688 GHz source of microwave.

Photocatalytic ammonia generation was performed in a 160 mL quartz reactor at ambient temperature and pressure. Catalyst with 100 mg mass was added to 80 mL of high-purity deionized water and 20 mL of methanol. Nitrogen gas was constantly bubbled into the system with a 60 mL min$^{-1}$ flow rate, while keeping the temperature constant at 298 K by a cold-water bath. The mixture under the dark condition was continuously stirred for 1 h to achieve an adsorption-desorption equilibrium of reactants on the catalyst surface. Subsequently, a 300 W full arc of xenon lamp with an intensity of 15 W m$^{-2}$ was used to illuminate the reaction solution. At certain intervals of 30 min, 2 mL of the liquid phase was sampled, filtered through 0.22 µm polytetrafluoroethylene hydrophilic filters, and analyzed by Nessler's reagent technique to measure the $NH_4^+$ concentration. Two types of control or blank experiments were conducted: (i) with the catalyst in the dark, and (ii) without the catalyst under light irradiation. The stability of the photocatalysts for the reactions was assessed by XRD measurement after the photocatalytic tests.

Fig. 1 displays the crystal structures of the three synthesized HEOs, as determined by (a) XRD patterns (including experimental measurements and calculated diffraction patterns), (b) experimental Raman spectroscopy, and computational modelling using special quasi-random structures (SQS). The Hf-containing oxide exhibits a dual-phase structure, consisting of nearly equal proportions of monoclinic (*A*2/*m*) and orthorhombic (*Ima*2) phases. The Ga-containing oxide has two phases, a monoclinic (*C*2/*m*) phase and an orthorhombic (*Pbcn*) phase, with the orthorhombic phase dominating 88 wt% of the structure. The Zn-containing oxide is found to have only a monoclinic (*P*2/*c*) structure. The model crystal structures are presented in Fig. 1 for (c) the monoclinic phase of the Hf-containing HEO, (d) the orthorhombic phase of the Hf-containing HEO, (e) the monoclinic phase of the Ga-containing HEO, (f) the orthorhombic phase of the Hf-containing HEO, and (g) the monoclinic phase of the Zn-containing HEO. It should be noted that the accuracy of these models was confirmed by comparing their calculated powder diffraction profiles with XRD patterns in Fig. 1(a).

To examine the uniformity of elements in three HEOs, SEM-EDS was performed. The SEM-EDS mappings in Fig. 2(a-c) suggest a bimodal spreading of elements in Hf- and Ga-containing HEOs, but the distribution of elements is reasonably homogenous in the Zn-containing HEO. These elemental distributions are consistent with the dual-phase structure of Hf- and Ga-containing HEOs and the single-phase structure of the Zn-containing HEO. The high-resolution TEM micrographs and relevant FFT diffractograms in Fig. 2(d-g) verify the crystal structures reported by XRD: (d) monoclinic and (e) orthorhombic for Hf-containing HEO, (f) monoclinic and (g) orthorhombic for Ga-containing HEO, and (h) monoclinic for Zn-containing HEO. It is noticed from the SEM images that the particle sizes of the three HEOs are rather large in the micrometer range. The surface areas measured by either SEM or BET are 0.70, 0.48 and 0.36 $m^2/g$ for Hf-, Ga- and Zn-containing HEOs, respectively. The rather large particle size and small surface area result from the synthesis by high-pressure torsion and calcination at an elevated temperature.

The optical properties measured using UV-Vis light absorption are shown in Fig. 3(a). All the HEOs present light absorption in both the UV and visible ranges. However, the Hf-containing oxide with $d^0$ electronic configuration represents lower light absorbance compared to the Ga- and Zn-containing HEOs with mixed $d^0$-$d^{10}$ cationic configurations. Kubelka-Munk analysis suggests bandgaps of 3.2, 2.5 and 2.5 eV for Hf-, Ga- and Zn-containing HEOs. Since a narrower bandgap generally leads to enhanced photocatalytic activities, the HEOs with mixed electronic configurations are considered more promising. Additionally, the HEOs with mixed cationic configuration show visible-light absorbance and another gap at 1.8 eV due to the formation of defects, an issue that was reported in various materials synthesized by high-pressure torsion [32-34]. Low bandgap energies and aligning the light absorbance with the visible light region are desirable characteristics for advanced photocatalysts [35-38]. As displayed in Fig. 3(b), the band structure of these three HEOs satisfies the threshold required for the $N_2$/$NH_3$ reduction reaction, which has a chemical potential of -0.05 eV vs. NHE (normal hydrogen electrode) [39].

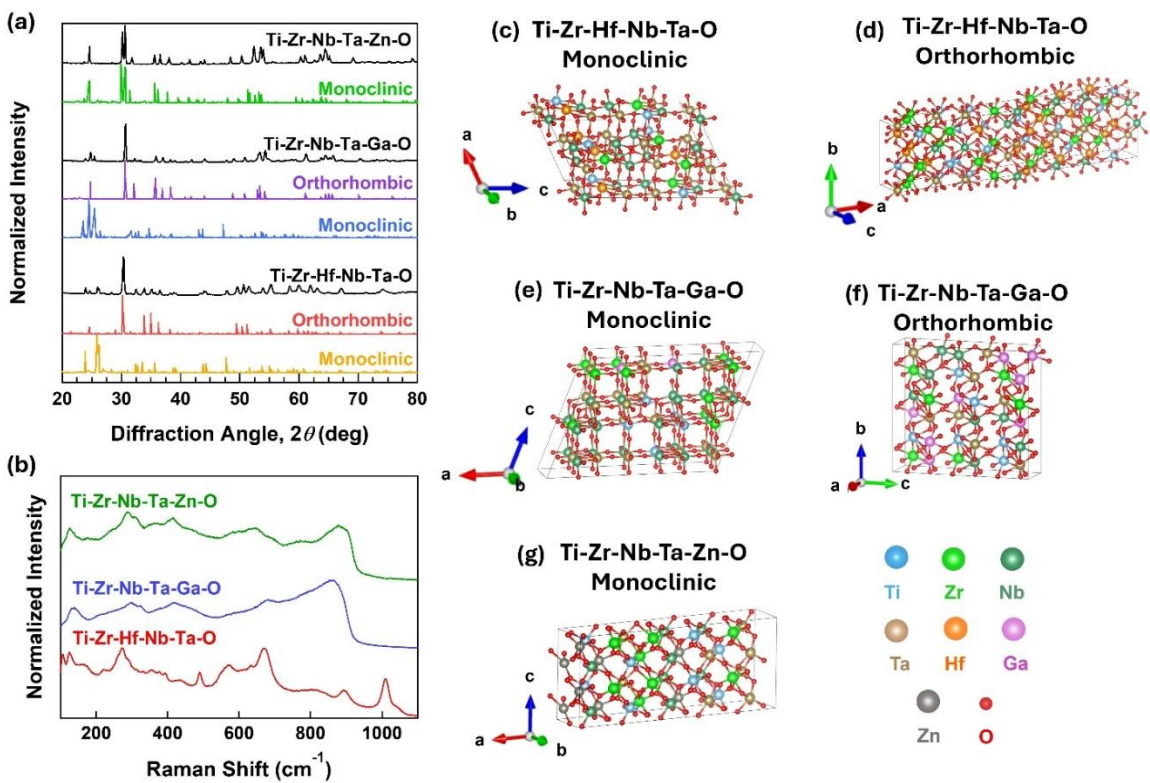

Figure 1. Crystal structures and structural modelling of high-entropy oxides Ti-Zr-Nb-Ta-X-O (X: Hf, Ga and Zn). (a) Experimental XRD profiles compared with calculated ones and (b) Raman spectroscopy profiles for three oxides. Modeled structures of (c) monoclinic and (d) orthorhombic phases of Hf-containing oxide, (e) monoclinic and (f) orthorhombic phases of Ga-containing oxide, and (g) monoclinic phase of Zn-containing oxide.

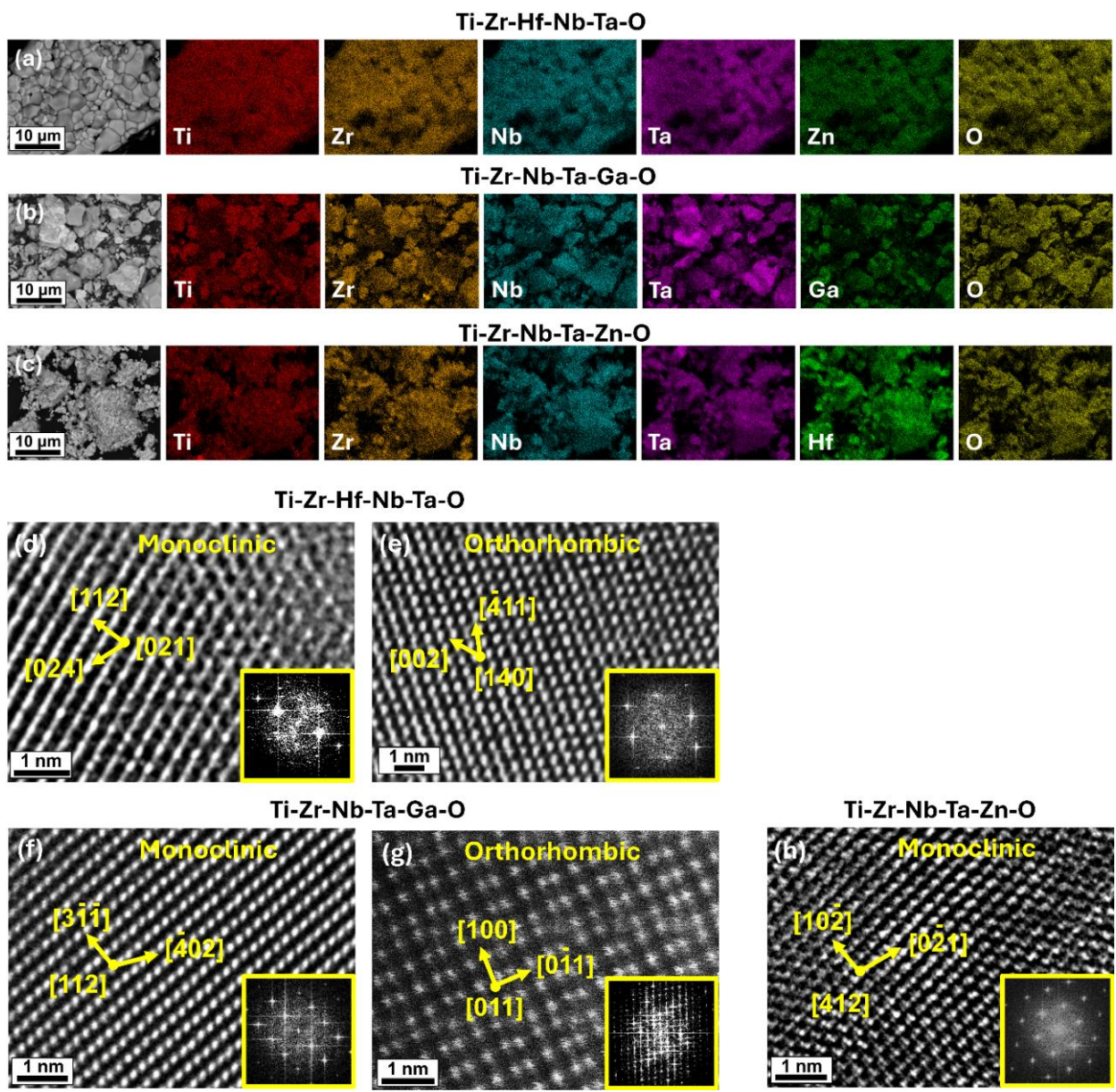

Figure 2. Elemental distribution and nanostructure of high-entropy oxides Ti-Zr-Nb-Ta-X-O (X: Hf, Ga and Zn). SEM images and corresponding EDS mappings of (a) Hf-, (b) Ga- and (c) Zn-containing oxides. High-resolution TEM images of (d) monoclinic and (e) orthorhombic phases of Hf-containing oxide, (f) monoclinic and (g) orthorhombic phases of Ga-containing oxide, and (h) monoclinic phase of Zn-containing oxide.

Charge carrier dynamics were investigated through steady-state PL and PL decay measurements, as illustrated in Fig. 3(c) and 3(d), respectively. The Hf-containing oxide possesses a strong steady-state PL peak at 552 nm, which can be attributed to irradiative recombination of charge carriers on defects. However, the introduction of mixed $d^0$-$d^{10}$ configurations in Ga- and Zn-containing oxides significantly reduces the PL intensity, indicating their lower electron-hole recombination compared to the HEO with only $d^0$ configuration. This decrease in radiative recombination is favorable for enhancing photocatalytic performance [40]. Furthermore, time-resolved PL decay measurements in Fig. 3(d) provide quantitative evidence for the lifetimes of photogenerated carriers. The decay profiles in Fig. 3(d) were analyzed using an exponential model.

$$I(t) = A_1 \exp(-t/\tau) \tag{1}$$

In this equation, $I(t)$ represents the PL intensity, $A_1$ is the amplitude parameter, $t$ represents the time after the excitation pulse and $\tau$ is the carrier lifetime. The fitting of PL-decay data results in lifetimes of 9.5, 11.6 and 27.0 ns for Hf-, Ga- and Zn-containing HEOs, respectively. The prolonged carrier lifetimes observed in the HEOs with mixed $d^0$-$d^{10}$ configuration indicate the persistence of photogenerated carriers, which is desirable for photocatalytic applications.

Fig. 3(e) and 3(f) present the photocurrent responses and EPR spectra of the synthesized HEOs, respectively, to provide insight into charge separation and defect characteristics. The photocurrent measurements in Fig. 3(e) indicate that the Ga- and Zn-containing HEOs, which incorporate a mixed $d^0$-$d^{10}$ electronic configuration, generate significantly higher photocurrents under chopped light illumination compared to the Hf-containing HEO with only $d^0$ cations. This result confirms the superior charge separation and mobility in the HEOs with mixed electronic configurations. The EPR spectra in Fig. 3(e) reveal stronger signals for oxygen vacancies in the Ga- and Zn-containing oxides, whereas the Hf-based HEO shows no visible oxygen vacancy signal. These findings suggest that good charge dynamics and the presence of oxygen vacancies in $d^0$-$d^{10}$ could result in enhanced photocatalytic activity compared to the HEO with only a $d^0$ configuration.

Fig. 4(a) presents the photocatalytic ammonia production activity over three HEOs. The photocatalytic tests were conducted for 3 h, and the amounts of ammonia production were measured every 0.5 h by Nessler's reagent colorimetric approach. While no ammonia is detected in the absence of a photocatalyst or without illumination, ammonia is formed by photocatalysis. After 3 h of light illumination, the generated ammonia amounts are 1340, 2742 and 3041 mg L$^{-1}$ m$^{-2}$ for Hf-, Ga- and Zn-containing HEOs, respectively. Fig. 4(b) compares the activity of Zn-containing HEO, as the most active catalyst with a mixed $d^0$-$d^{10}$ configuration, with the activity of a mixture of binary oxides. The activity of the HEO is considerably better than binary oxides, indicating high-entropy materials as promising active catalysts for photocatalytic ammonia production. Fig. 4(c) exhibits XRD spectra of all three photocatalysts applied in this study, both before and after the photocatalytic experiments. It is confirmed that the crystal structures remain unchanged during photocatalytic reactions, indicating the good stability of the photocatalysts.

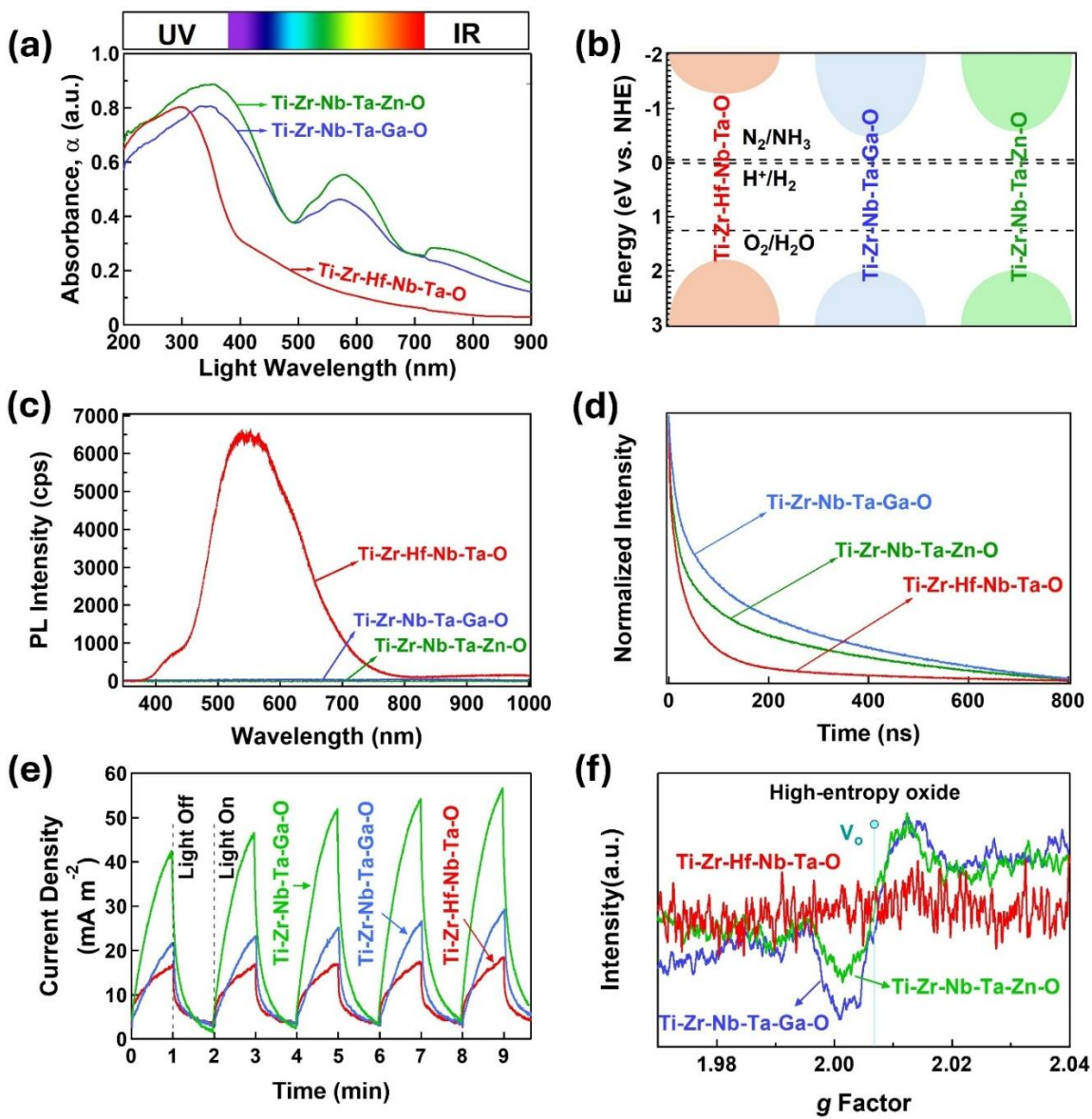

Figure 3. Comparison of light absorbance and recombination dynamics of high-entropy oxides Ti-Zr-Nb-Ta-X-O (X: Hf, Ga and Zn). (a) Light absorbance, (b) electronic band configuration, (c) steady-state PL spectroscopy, (d) PL decay spectroscopy, (e) photocurrent generation, and (f) EPR spectroscopy for three oxides.

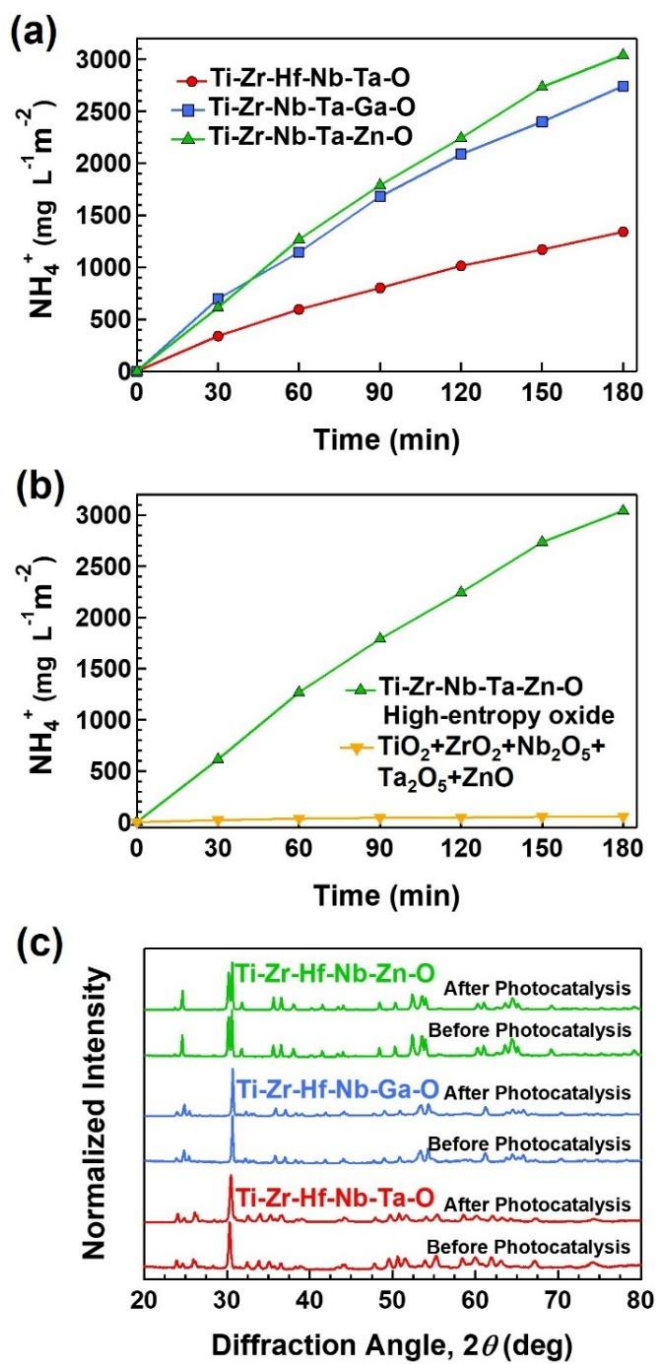

Figure 4. Photocatalytic activity and stability of high-entropy oxides Ti-Zr-Nb-Ta-X-O (X: Hf, Ga and Zn) for ammonia synthesis. (a) Ammonia production versus irradiation time for (a) three oxides, and (b) Zn-containing oxide compared to corresponding binary oxides. (c) XRD patterns of three oxides before and after photocatalysis.

Although this investigation presents the first successful application of HEOs for photocatalytic ammonia synthesis, three questions should be still discussed: (i) the pathway for photocatalytic ammonia synthesis over HEOs, (ii) the causes for higher photocatalytic activity of HEOs with mixed $d^0$-$d^{10}$ cationic configuration, and (iii) comparison of photocatalytic ammonia production over the HEOs with other reported photocatalyst systems.

Regarding the first question, the photocatalytic ammonia production process over the HEOs should follow a multi-step reduction of nitrogen gas molecules, driven by photogenerated electrons and protons, driven by water oxidation [41–44]. Firstly, upon illumination, the HEOs absorb photons and generate electron-hole pairs. Secondly, the photogenerated electrons migrate to surface-active sites, where they reduce adsorbed $N_2$ molecules into $NH_3$, while the holes contribute to the oxidation of water or methanol. Thirdly, the resulting $NH_3$ molecules desorb from the high-entropy oxide surface, completing the catalytic cycle. As depicted in Fig. 3(b), the conduction band configuration of all investigated high-entropy oxides is more negative compared to the chemical potential of the $N_2/NH_3$ redox reaction, confirming their thermodynamic feasibility for $N_2$ reduction. During the first proton-coupled electron movement, the nitrogen gas molecules absorbed on the surface of the catalyst, in the *N≡N form, receive an electron ($e^-$) of the catalyst and a proton ($H^+$) of the reacting environment, resulting in the formation of an adsorbed chemical specie (*N=NH) as follows [41,45,46].

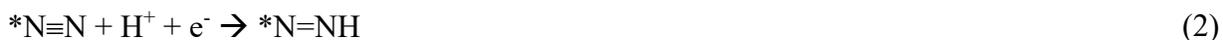

*N≡N + $H^+$ + $e^-$ → *N=NH                                                                 (2)

Density functional theory calculations suggested five routes for the subsequent transfer of $H^+/e^-$ (Fig. 5) [47], although two of them are the most recognized processes to reduce $N_2$ to $NH_3$. These two routes, which are referred to as distal and alternating mechanism routes, are illustrated in Fig. 5. The distal mechanism (route 1 in Fig. 5) suggests that $H^+/e^-$ pairs bind to only one of the nitrogen atoms of $N_2$ to create a terminal nitride intermediate. This binding releases the first ammonia molecules and leaves another nitrogen atom, which is then transformed to form the second ammonia molecule. The alternating mechanism assumes that the pairs of $H^+/e^-$ are bound to both atoms of nitrogen gas (route 2 in Fig. 5), where further binding of $H^+/e^-$ to these atoms finally leads to $NH_3$ production.

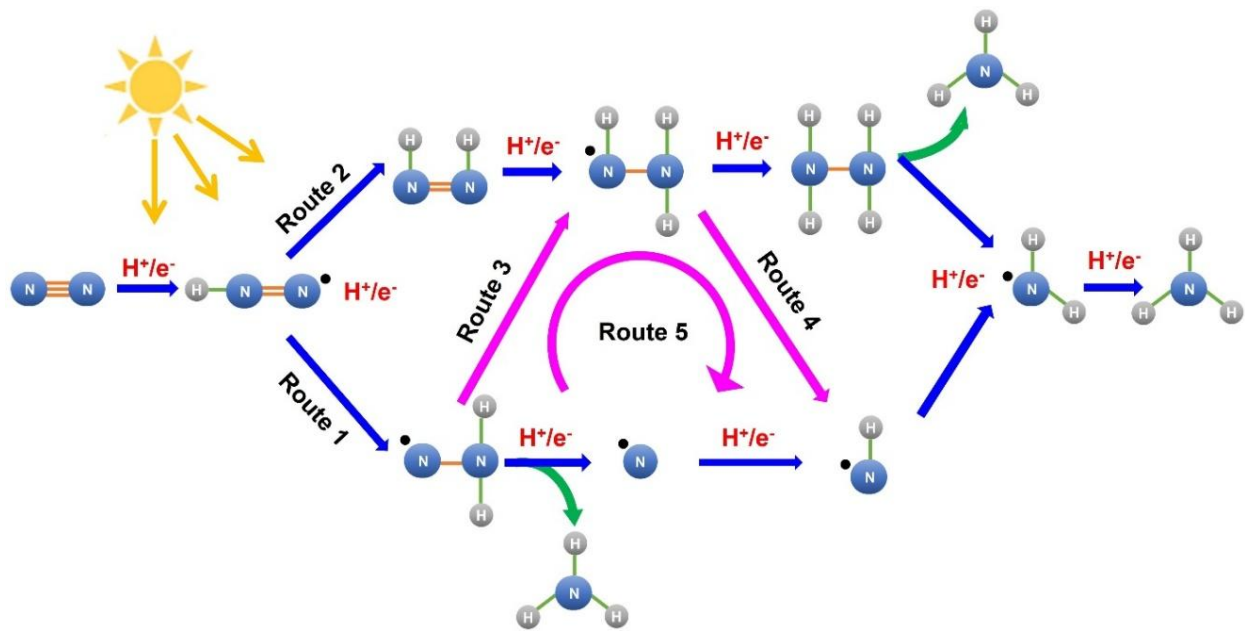

**Figure 5.** Proposed pathways for $N_2$ reduction to $NH_3$. Routes 1 and 2 are referred to as distal and alternating mechanisms.

Regarding the second question, the role of $d^0$ cations in photocatalysis is well established, particularly in applications such as hydrogen production, $CO_2$ conversion and plastic transformation [31]. The Hf-containing oxide, which contains only $d^0$ cations, demonstrates a notable activity for $NH_3$ production, which indeed suggests a new research direction for high-entropy photocatalysts. However, when hafnium is substituted with a $d^{10}$ element such as gallium or zinc, the photocatalytic activity shows an enhancement in ammonia production (Fig. 4(a)). This improvement can be ascribed to the combined geometric-electronic effects induced by the incorporation of cations with diverse electronic configurations. Structurally, the existence of cations with varying atomic sizes introduces lattice distortion, which, in turn, affects octahedral and tetrahedral coordination. These distortions influence ionic features, electron transport and overall photocatalytic performance by generating dipole moments within the material [48-51]. Yuan et al. demonstrated that introducing lattice distortion can effectively induce oxygen vacancies and alter the electronic band structure, thereby improving the photocatalytic properties of materials. These modifications enhance charge carrier mobility, reduce recombination loss, and increase electron-driven reduction potential. Additionally, the elevated concentration of oxygen vacancies plays a key role in facilitating $N_2$ molecule adsorption and activation for photocatalytic

N$_2$ fixation [51]. From an electronic perspective, the d$^0$ elements primarily act as electron acceptors, whereas d$^{10}$ elements can also serve as electron donors. Due to their lower electronegativity, d$^0$ cations exhibit strong binding interactions with adsorbed N$_2$ or H$_2$O molecules [52]. Meanwhile, the inclusion of d$^{10}$ elements contributes to the band structure via the sp orbitals, enhancing charge carrier mobility and promoting charge separation [50]. Additionally, the enhanced generation of oxygen vacancies in Ga- and Zn- containing HEOs, as evidenced by EPR analysis (Fig. 3(f)), suggests that the incorporation of d$^{10}$ cations promotes the generation of surface defects, which are widely recognized as active sites for N$_2$ absorption and activation, thereby further contributing to the improved photocatalytic performance. As a result, the mixed d$^0$-d$^{10}$ electronic configuration creates a heterogeneous chemical environment that facilitates efficient charge transfer and strengthens the adsorption of N$_2$ and H$_2$O. This combination leads to superior photocatalytic properties, including enhanced light absorption (Fig. 3(a)), a reduced bandgap (Fig. 3(b)), lower electron-hole recombination rates (Fig. 3(c)) and an extended charge carrier lifetime (Fig. 3(d)). Subsequently, the mixed electronic configuration enhances the photocatalytic performance compared to the d$^0$ electronic configuration.

Regarding the third question, Table 1 compares photocatalytic ammonia synthesis using the HEOs with previously reported catalyst groups. Among various photocatalytic systems, the HEOs demonstrate notable activity and efficiency in ammonia production, suggesting that they are promising candidates for photocatalytic ammonia synthesis. This enhanced performance is because of the exclusive characteristics of high-entropy materials, including large diversity in their composition and disordering of their structure, which contribute to photocatalytic efficiencies superior to traditional catalysts [24,53]. The co-presence of various cations in the lattice generates a dynamic and complex catalytic environment, fostering synergistic interactions that enhance catalytic efficiency. Additionally, the inherent structural disorder in HEOs facilitates efficient charge separation and migration, resulting in improved charge carrier generation, reduced recombination losses, and ultimately, higher overall photocatalytic performance. Because of their strong structure and high configurational entropy, HEOs have remarkable thermal and chemical stability in addition to their catalytic effectiveness [24]. This stability prevents material degradation and efficiency decline under catalytic reaction conditions, ensuring long-term durability. Given these advantages, HEOs emerge as a promising family of photocatalysts for ammonia synthesis. Future studies are expected to focus on determining the apparent quantum efficiency of HEOs for ammonia production, optimizing compositions of HEOs and synthesizing them with processes other than high-pressure torsion, which is more appropriate for laboratory-scale synthesis of high-entropy alloys [63] and their ceramics [64].

Table 1. Comparison of ammonia production by photocatalytic process from high-entropy oxides Ti-Zr-Nb-Ta-X-O (X: Ga and Zn) with reported data for some catalysts.

| Catalysts | Ammonia generation rate (mg L$^{-1}$ h$^{-1}$ g$^{-1}$) | Ref. |
|---|---|---|
| Ni$_2$P/Cd$_{0.5}$Zn$_{0.5}$S | 1.73 | [54] |
| MoS$_2$/CdS | 1.87 | [55] |
| CdS@Ti$_3$C$_2$-15 | 4.98 | [56] |
| Fe$_2$O$_3$/g-C$_3$N$_4$ | 47.90 | [57] |
| W$_{18}$O$_{49}$/g-C$_3$N$_4$ | 0.52 | [58] |
| CuCr/LDH | 3.14 | [59] |
| g-C$_3$N$_4$/Ag$_2$CO$_3$ | 1.10 | [60] |
| g-C$_3$N$_4$/Zn$_{0.11}$Sn$_{0.12}$Cd$_{0.88}$S$_{1.12}$ | 1.51 | [14] |
| Zn$_{0.1}$Sn$_{0.1}$Cd$_{0.8}$S | 0.05 | [61] |
| 0D-2D Bi$_4$O$_5$Br$_2$/ZIF-8 | 0.28 | [62] |
| Ti-Zr-Ta-Nb-Ga-O | 32.91 | This work |
| Ti-Zr-Ta-Nb-Zn-O | 36.50 | This work |

In conclusion, this study highlights the promise of high-entropy photocatalysts as new candidates for ammonia production from atmospheric nitrogen. The high structural disorder degree and compositional diversity in these photocatalysts promote synergistic effects, resulting in enhanced ammonia production compared to traditional binary oxides. Moreover, the combination of d$^0$ and d$^{10}$ electronic configuration elements in these oxides enhances photocatalytic performance through improved charge separation, increased light absorption, decreased charge carrier recombination and strengthened adsorption of reactant molecules. These findings suggest the potential of high-entropy ceramics for sustainable ammonia synthesis, opening the door for further research and advancements.

**CRediT Authorship Contribution Statement**

All authors: Conceptualization, Methodology, Investigation, Validation, Writing – review & editing.

**Declaration of Competing Interest**

The authors declare no competing financial interests or personal relationships that could affect the results presented in this article.

**Acknowledgments**

The author JHJ is grateful to the Q-Energy Innovator Fellowship of Kyushu University for a scholarship. This work is supported in part by Mitsui Chemicals, Inc. of Japan, and in part by the Japan Science and Technology Agency through the Establishment of University Fellowships Towards the Creation of Science Technology Innovation (JPMJFS2132) and ASPIRE Project (JPMJAP2332).